\documentclass[apj]{emulateapj}
\usepackage{apjfonts}

\usepackage{amsmath}
\usepackage{natbib}

\submitted{Accepted by the Astrophysical Journal Supplements}

\usepackage{graphicx}
\usepackage[figuresright]{rotating} 
\usepackage{float}       
\usepackage{afterpage}   

\newcommand {\GRB}{GRB~041219a}

\newcommand {\GR}{$\gamma$-ray}
\newcommand{\integral}{\emph{INTEGRAL}}
\newcommand{\cgro}{\emph{CGRO}}
\newcommand{\wsim}{\ensuremath{\sim}}
\newcommand{\rxte}{\emph{RXTE}}

\begin{document}


\title{Search for polarization from the prompt gamma-ray emission of
GRB~041219a with SPI on \emph{INTEGRAL}}

\author{E. Kalemci\altaffilmark{1},
        S. E. Boggs\altaffilmark{2,3},
        C. Kouveliotou\altaffilmark{4,5},
        M. Finger\altaffilmark{4,6},
        M. G. Baring\altaffilmark{7},
}

\altaffiltext{1}{Sabanc\i\ University, Orhanl\i
-Tuzla, \.Istanbul, 34956, Turkey.}

\altaffiltext{2}{Space Sciences Laboratory, 7 Gauss Way, University of
California, Berkeley, CA, 94720-7450, USA.}

\altaffiltext{3}{Department of Physics, University of California,
366 Le Conte Hall, Berkeley, CA, 94720-7300, USA.}

\altaffiltext{4}{National Space and Technology Center, 320 Sparkman
Drive, Huntsville, AL, 35805, USA.}

\altaffiltext{5}{NASA Marshall Space Flight Center, SD-50,
Huntsville, AL, 35812, USA.}

\altaffiltext{6}{Universities Space Research Association}

\altaffiltext{7}{Department of Physics and Astronomy, MS-108, Rice
University, P.O. Box 1892, Houston, TX 77251}


\begin{abstract}

Measuring the polarization of the prompt \GR\ emission from GRBs can
significantly improve our understanding of both the GRB emission mechanisms,
as well as the underlying engine driving the explosion. We searched for
polarization in the prompt \GR\ emission of \GRB\ with the  SPI instrument on
\integral. Using multiple-detector coincidence events in the 100--350 keV
energy band, our analysis yields a polarization fraction from this GRB of
98 $\pm$ 33\%.  Statistically, we cannot claim a polarization detection from 
this source. Moreover, different event selection criteria lead to even less 
significant polarization fractions, e.g. lower polarization fractions are 
obtained when higher energies are included in the analysis. We cannot strongly 
rule out the possibility that the measured modulation is dominated by 
instrumental systematics. Therefore, SPI observations of \GRB\ do not 
significantly constrain GRB models. However, this measurement demonstrates the 
capability of SPI to measure polarization, and the techniques developed for 
this analysis.

\end{abstract}




\section{Introduction}\label{sec:intro}

Despite the extensive work in recent years on GRB afterglows, the nature of
 the central driver that powers the burst and the prompt \GR\ emission
mechanism remain enigmatic, as the physics of the afterglow is
insensitive to the nature of the progenitor once a relativistic
fireball is formed. There has been different suggestions for the
mechanism that powers the GRB central engine. In the models invoking
merging neutron stars and 'collapsars'
\citep{Woosley93,Paczynski98,MacFadyen99}, hydrodynamically
dominated outflows (jets) transport the bulk GRB kinetic energy.
Alternatively, Poynting-flux may be the driver for the transport of
energy to large distances \citep{Lyutikov03}.  Synchrotron radiation
has traditionally been the favored emission mechanism of the prompt
\GR\ emission \citep{Meszaros94,Tavani96, Dermer99,Lloyd00}, though
competing Compton upscattering and synchrotron-self Compton models
have been put forward
\citep{Liang97,Meszaros94,Chiang99,Sari01,Zhang01}; reviews of GRB
models can be found in \cite{Piran99} and \cite{Meszaros01}. In
terms of polarization modeling, synchrotron radiation is naturally a
strong candidate \citep{Coburn03,Granot03}, but a portion of the
polarized photon signal may also be Compton up-scattered
\citep{Eichler03}. A definite measurement of polarization properties
from the prompt emission of GRBs will probe their anisotropy or magnetic field
geometry, and thereby help determine the nature of the central engine and the 
\GR\ emission mechanism.

\begin{figure*}[t]
\plotone{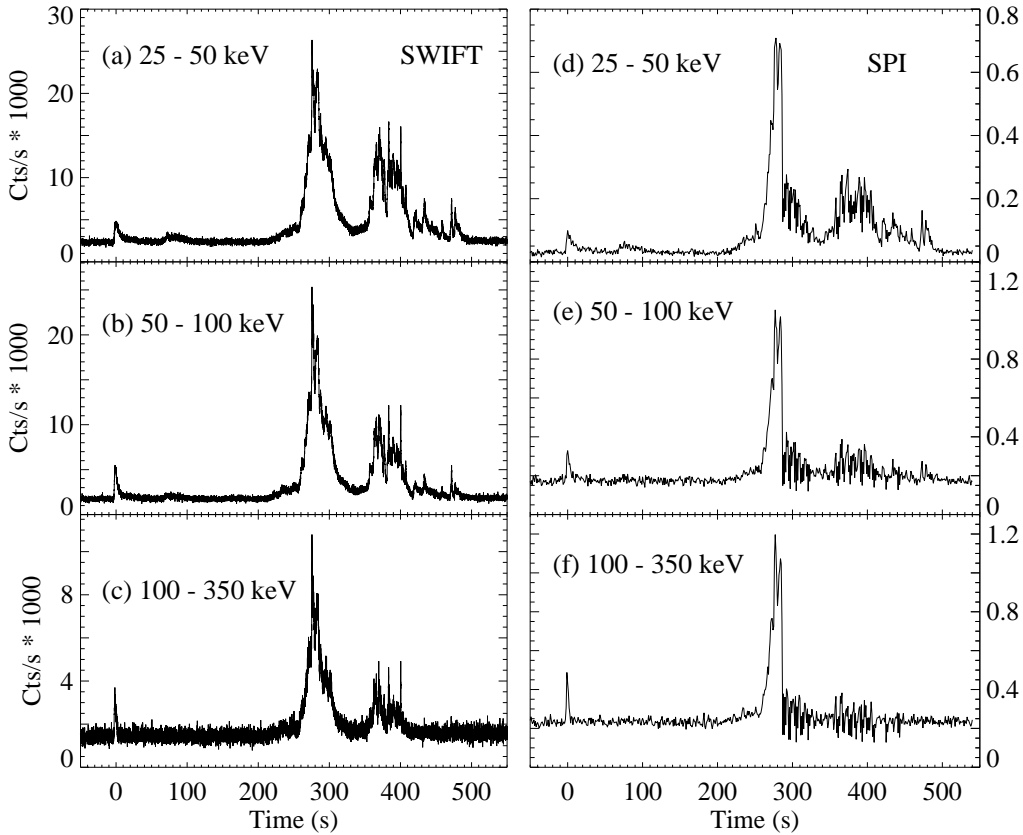}
\caption{\label{fig:spiswift}
The \emph{SWIFT} light curves of \GRB\ in (a) 25--50 keV, (b) 50--100 keV and
(c) 100-350 keV bands \citep{Barthelmy04GCN}, and the SPI light curves for the
singles in the same bands in (d), (e), and (f) respectively. Times are relative
 to the peak of the precursor.
}
\end{figure*}


The first detection of the linear polarization from the prompt \GR\
emission of a GRB indicated a very high polarization fraction of 80
$\pm$ 20 \% \citep{Coburn03}. For this measurement, Reuven Ramaty
High Energy Solar Spectroscopic Imager \citep[\emph{RHESSI}, ][]{Lin02}
data of GRB~021206 were used. The measurement demonstrated the
potential for measuring polarization using Compton-scattered events
between multiple detectors. Using \emph{RHESSI}, the same method is 
used later to measure the polarization fraction of two X-class solar flares 
\citep{Boggs06}. 

The large polarization fraction obtained by \cite{Coburn03} resulted
in a series of theoretical work on \GR\ polarization in GRBs and
Poynting dominated flows
\citep{Lyutikov03,Nakar03,Eichler03,Granot03,Dai04,Lazzati04}.
However, independent analyses of the \emph{RHESSI} data by other
groups were not able to confirm this result at the same level of
significance \citep{Wigger04,Rutledge04}, so that the degree of
polarization for GRB~021206 remains uncertain. Clearly, more
measurements, using different instruments and techniques, are
required.  Recently, using the BATSE instrument on \cgro,
\cite{Willis05} provided evidence for large polarization fractions
for two bursts, GRB~930131 ($\Pi \; >$ 35\%) and GRB~960924 ($\Pi \;
>$ 50\%), without strongly constraining the upper limits. In their
work, the mass model of BATSE, along with a mass model of the
earth's atmosphere were used, and the polarization fraction was
determined by analyzing the angular distribution of photons that are
scattered through the earth's atmosphere. SPI \citep[Spectrometer on
\integral,][]{Vedrenne03} and IBIS \citep[Imager on Board the
\integral\ Satellite,][]{Ubertini03} instruments on \integral\ can
also measure the polarization fraction and angle of a source using
the coincidence events between detector pairs
\citep{Lei97,Kalemci04_int}, similar to the method employed by
\cite{Coburn03,Boggs06} with RHESSI.

 In this letter, we discuss methods to measure polarization using one of
the instruments on \integral, SPI, and apply these methods to
measure the polarization properties of \GRB, a bright and a long
(about 450 s) GRB which was detected with the \integral\ Burst Alert 
System (IBAS) and with  ISGRI on December 19 at 01:43 UT. The burst
is in the fully coded field of view of both the ISGRI and the SPI,
and is \wsim 3$^\circ$ off the X-axis, and 155$^\circ$ in azimuth from
the Y-axis \footnote{In this work, the azimuthal angles are defined in a plane 
perpendicular to the SPI pointing X-axis, and measured with respect to the 
SPI Y-axis towards the SPI Z-axis.}. The ISGRI coordinates are
reported as RA = 6.1075$^{\circ}$, and DEC=+62.8349$^{\circ}$ with
an uncertainty of 2$^{\prime}$ \citep{Gotz04GCN}. The brightest part
of the burst saturated the available telemetry of \integral. The
long duration and brightness allowed for multi-wavelength campaigns
for this GRB \citep{Blake04GCN, Blake04GCN2,
Soderberg04GCN,Sonoda04GCN}. The infrared counterpart location is
given as RA=6.1153$^{\circ}$, and DEC=62.8426$^{\circ}$
\citep{Blake04GCN2}. The burst was also detected with
\emph{SWIFT}-BAT \citep{Barthelmy04GCN,Fenimore04GCN}. A comprehensive 
spectral and temporal analysis of the burst with SPI, \emph{SWIFT}-BAT, and 
the \rxte\ ASM is given in \cite{McBreen06}. The
\emph{SWIFT}-BAT and SPI (singles) light curves are shown in
Fig.~\ref{fig:spiswift}. These light curves indicate that the
spectrum softens as the burst progresses, and also show a precursor
\wsim 250 s before the main peak.

\section{Analysis}\label{sec:obs}

\subsection{SPI and \GR\ polarization}\label{subsec:spipol}

SPI is a coded-aperture telescope using an array of 19 cooled
germanium detectors for high-resolution spectroscopy
\citep{Vedrenne03}. It works in 20 keV -- 8 MeV band, and has an
energy resolution of \wsim 2 keV below 500 keV. The fully coded
field of view is 16$^{\circ}$, and the angular resolution is \wsim
3$^{\circ}$. At the time of the observation, 17 detectors were
active due to the failures of Detectors 2 and 17 in orbit. If a
photon deposits all of its energy into one detector, SPI records
this as a single event. If a photon interacts through Compton
scatterings with energy deposits in more than one detector, the
detector and channel information for each interaction are saved into
a multiple event (ME).  Even though SPI is not primarily designed for 
polarization measurements, these ME data are inherently sensitive to 
polarization as linearly polarized gamma-rays preferentially scatter in 
azimuthal directions perpendicular to their electric polarization vector 
\citep{Kalemci04_int}.

The two main parameters that determine the sensitivity of a multi-detector
 instrument to gamma-ray polarization are the effective area to the
multiple-detector scatter events, and the average value of the polarimetric
modulation factor $Q$, which is the maximum variation in the azimuthal
scattering probability for polarized photons \citep{Novick75,Lei97}. This
factor is determined by the scattering cross sections,
\begin{equation}\label{eq:Q}
Q=\dfrac{(d\sigma_{\perp} - d\sigma_{\parallel})}{(d\sigma_{\perp} + d\sigma_{\parallel})} ,
\end{equation}
where d$\sigma_{\perp}$, d$\sigma_{\parallel}$ are the
Klein-Nishina differential cross-sections for Compton scattering perpendicular
and parallel to the polarization direction, respectively, which are functions
of the incident photon energy and the Compton scatter angle between the
incident photon direction and the scattered photon direction.  For a source
count rate of $S$, and fractional polarization of $\Pi_{s}$, the expected
azimuthal scattering angle ($\phi$) distribution is
\begin{equation}\label{eq:pol}
{\dfrac{dS}{d\phi}} = {\dfrac{S}{2\pi}}[1-Q \, \Pi_{s} \cos 2(\phi-\eta)] .
\end{equation}
Therefore, the ``signature'' of polarization is a 180$^\circ$-periodic
modulation in the distribution of azimuthal scattering angles, with a minimum
at the polarization angle $\eta$.

\subsection{\GRB\ SPI data}

 A first look at the light curve showed that the SPI
data for this GRB were affected by the telemetry saturation problems
which also affected the IBIS data \citep{Gotz04GCN}.
Fig.~\ref{fig:spiswift} shows the SPI light curves (right block) for
singles for the sum of all detectors. Here we define ``singles''
-- all single detector events -- as the sum of all SE and PSD events in
the Integral Science Data Center (ISDC) format. As the flux peaks, a
sudden drop occurs in the count rate, which was not observed in the
\emph{SWIFT} light curves (left block).

\begin{figure}[t]
\plotone{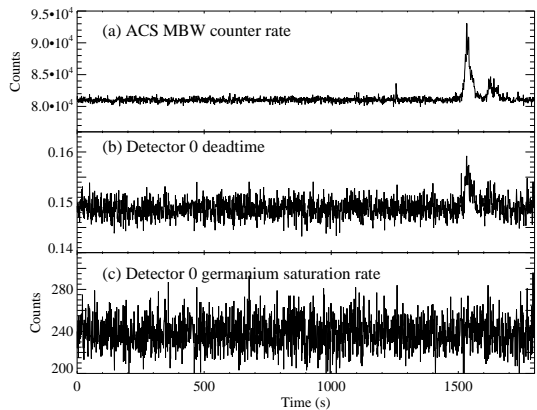}
\caption{\label{fig:hsk}
(a) The ACS MBW counter rate, (b) Detector 0 dead-time, (c) Detector 0
germanium saturation rate. Time 0 is the beginning of pointing.
}
\end{figure}


We inspected several housekeeping parameters to verify that the
origin of the problem is not something other than telemetry saturation. The
anti-coincidence system rate, the dead-time, and the germanium saturation rate
for Detector 0 are shown in Fig.~\ref{fig:hsk}. Even though there is an
increase in dead-time, the increase is modest, and cannot account for the
dropouts in SPI light curves. The germanium saturation rate shows no
significant deviation from the norm that could cause a sudden decrease in the
count rate. We then compared the ``raw''  and the ``prp'' on-board time
(processed through a standard pipeline at the ISDC), and found that the prp
times have gaps that are approximately multiples of 0.125 s, indicating that
telemetry packets are missing (S. Schanne, private communication). Since the
time and duration of the gaps are known, an approximate light curve can be
reconstructed. The 100-500 keV (total energy) light curve of ME events,
corrected for effective dead-time due to the missing packets, is shown in
Fig.~\ref{fig:recon}. Characterizing this effective dead-time is important in
terms of determining the correct background rate for the regions with the
packet loss problem.

\subsection{MGEANT simulations}\label{subsec:mgeant}

To determine the polarization fraction for this GRB, we need to compare the
measured azimuthal scattering angle distribution to the expected distribution
for an unpolarized and a polarized source from this sky location. The only
method available for performing this comparison is with detailed
Monte-Carlo simulations. The response to a polarized source is characterized
by the polarimetric modulation factor, $Q$, discussed in
\S~\ref{subsec:spipol}. Since $Q$ is energy-dependent, it will depend on the
energy spectrum of the source. We therefore used simulations for two
purposes, to obtain the spectral parameters (\S~\ref{subsec:spispe}), and to
obtain the modulation factors.

The simulations are performed using MGEANT \citep{Sturner00}, which is a \GR\
instrument simulation package developed at NASA/GSFC. The MGEANT source code
allows several beam geometries and spectra to be specified at compile time.
 A highly detailed SPI mass model is  used as an input to MGEANT. 
In order to have a complete response, the mass model of the rest of the
spacecraft \citep{Ferguson03} is also included. This mass model is the same as
the mass model used to create SPI response matrices with MGEANT. 
More information on MGEANT and the complete mass model we used can be found 
in \cite{Sturner03}.

\subsection{\GRB\ spectrum}\label{subsec:spispe}


\begin{figure}[t]
\plotone{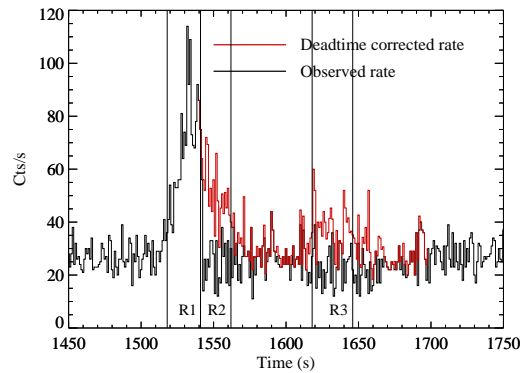}
\caption{\label{fig:recon}
The observed (black histogram) and the reconstructed (red histogram) light
curve of ME events in 100--500 keV band. The gaps are treated as dead-time.
The vertical solid lines separate R1, R2, R3 regions (see text).
}
\end{figure}

\begin{figure}[t]
\plotone{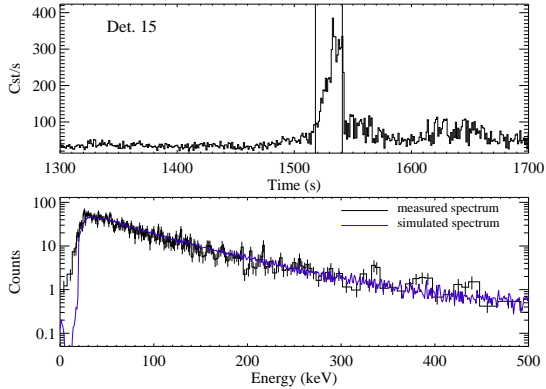} 
\caption{\label{fig:det15spe} The top
panel shows the singles light curve of Detector 15 in 30--490 keV
band. No dead-time correction is applied. The solid vertical lines
indicate the region (R1) for which the spectrum is extracted. The
bottom panel shows the measured (black histogram), and the simulated
(purple) spectrum. }
\end{figure}

Detailed and precise determination of the \GRB\ spectral parameters
is not necessary for this work as $Q$ is not strongly dependent on the exact 
spectral parameters. Therefore, a rough determination of the GRB spectrum is 
adequate for our study. To determine the spectral parameters, we first 
obtained the singles count spectrum from the region with no packet loss (R1 in 
Fig.~\ref{fig:recon}). We determined the background for each detector 
as follows: We took the data from the first 1000 s from the beginning 
of the pointing and obtained a spectrum. Next we applied two corrections. 
We fit the background light curve with a first order polynomial to take into
account a small (a few percent) and gradual increase towards the
GRB. Second, we found the live-times at the background region
($B_{livetime}$) and the GRB region ($GRB_{livetime}$), and
multiplied the background spectrum with
{$GRB_{livetime}$/$B_{livetime}$.} An example spectrum after
background subtraction is shown in Fig.~\ref{fig:det15spe}.


After the photons from the GRB event were isolated and spectra for
each detector were obtained, the next step is to reproduce these
spectra with simulations. To perform the simulations, we first
modified the original mass model such that the detectors 2 and 17
are not active. We obtained longitude and latitude of the spacecraft
axes using ``\emph{spipoint}'', and used the position of the
infrared counterpart for the GRB. We ran three simulations with the
Band Function \citep{Band93} spectrum using (1) $\alpha$=1.0,
$\beta$=2.4, $E_{br}$=170, (2) $\alpha$=1.0, $\beta$=2.0,
$E_{br}$=170, and (3) $\alpha$=1.0, $\beta$=2.0, $E_{br}$=200.

For simplicity, the mass model uses a single mass for all detectors,
even though in reality the mass of each detector is slightly
different. This mass distribution causes the largest
detector-to-detector variations in efficiency \citep{Sturner03}. The
simulated spectra were corrected for this effect. We also applied a
correction for dead-time for each detector. Apart from these
detector-dependent corrections, there are also detector independent
corrections regarding the photo-peak efficiencies and the mask
transmission. These were also applied as described in
\cite{Sturner03}. We found that the spectrum with these set of
parameters, $\alpha$=1.0, $\beta$=2.0, $E_{br}$=200, best describes
the data in R1. In Fig.~\ref{fig:det15spe}, we show the actual and
the simulated spectrum of Detector 15 as an example \footnote{These spectral
results were obtained before the publication of \citealt{McBreen06}. Even 
though band function parameters are different, the effect of small differences
 in the energy spectrum is not important for polarization measurements.}

\begin{figure}[t]
\plotone{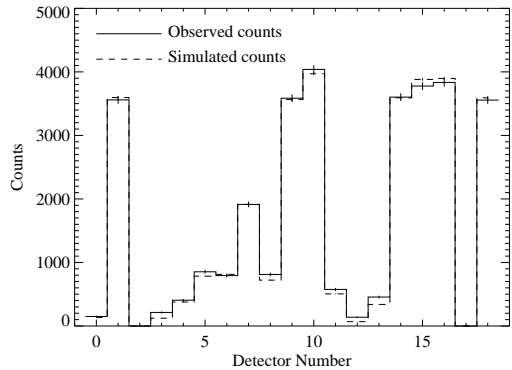}
 \caption{\label{fig:comp}
Comparison between the actual (solid histogram) and the simulated (dashed
histogram) number of counts for each detector for singles, in
30--490 keV band.
}
\end{figure}

We also checked the detector distribution of 30--490 keV band total singles
counts and compared it to the simulated distribution. The result is shown in
Fig.~\ref{fig:comp}. The simulation (shown with dashed lines) reproduced the
actual distribution well (within a few \%) for detectors not shadowed by
the mask. For the detectors under the shadow of the mask elements, the
discrepancy is larger if expressed in percentage. Note that the number of
events in these detectors are much less than the open detectors. Finally we
compared the total number of ME after cuts (see \S~\ref{sec:polme}) 
with the total number of ME from simulations. We obtained 534
ME from simulations compared to 543 events from the data, an agreement within
2\%.


Overall, the Band spectrum with $\alpha$=1.0, $\beta$=2.0, $E_{br}$=200.
represents the GRB in R1, and the simulation reflects the actual
detector to detector distribution.

\subsection{Modulation Factor}\label{subsec:Q}

The next step to measure polarization is to obtain $Q$ (see
Eq.~\ref{eq:Q}) by comparing the azimuthal scattering angle
distributions of non-polarized and 100\% polarized photons. Both can
be created using MGEANT simulations. Determining the azimuthal scattering angle
requires finding the direction of the photon as it scatters from one detector
to the other. SPI records the energies and the detectors in a ME. But the 
direction of the photon cannot be uniquely determined for all events. The 
conservation of energy and momentum in the Compton scattering process place 
limits on the energies deposited in each detector. Assuming full energy 
deposition in two detectors after a single Compton scattering:
\begin{equation}
  {{m_{e}c^{2}}\over{E_{1}}} - {{m_{e}c^{2}}\over{E_{1}+E_{2}}} = 1 - \rm{cos(\theta)}
\end{equation} 

where $E_1$ is the energy deposition in the first detector, and $E_2$ is 
the energy deposition in the second detector, and $m_{e}$ is the mass of the
electron. One can easily show that for relatively small total energies 
($E_1+E_2 \leq {{{m_{e}c^{2}}}\over{2}}$) $E_2$ is always 
greater than $E_1$. As the initial energy increases, the number of cases with
$E_2 < E_1$ increases, and finally at ${m_{e}c^{2}}$=511 keV, there is equal 
probability for either case. For the spectrum of GRB~041219, most of the 
photons Compton scattered from the low energy deposition detector to the high
energy deposition. Therefore we tag the direction of every photon as 
originating from the lower energy deposition detector. Even
though some of the interactions will be tagged incorrectly this way, the final
results should not be affected significantly due to the  180$^{\circ}$ symmetry
of the polarization modulations. 

On the other hand, MGEANT simulations provide more information than that of the
real data. First, in simulations, the interaction positions within the 
individual detectors are known. Second, for any incoming photon energy, the
direction of the photon is also known. We determined azimuthal scattering 
angle distributions for three cases; (a) using the actual interaction 
positions and directions determined by the simulation, (b) using the detector 
center-center angles (pixellation) and directions determined by the 
simulation, (c) using the center-center angles and directions determined using 
energy depositions. Cases (a) and (b) can only be calculated using the 
simulations, and (c) represents distribution for the actual data.


\begin{figure}[t]
\plotone{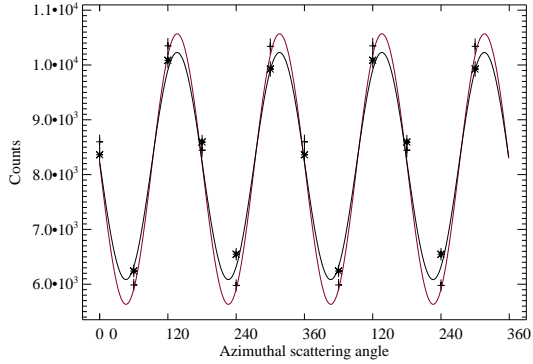}
\caption{\label{fig:mod1}
Simulated azimuthal scattering angle distribution of 100\% polarized photons
at 200 keV originating at the GRB position. The center-to-center pixellated
distribution (black fit) has lower modulation amplitude than the
distribution obtained using interaction positions within the detectors
(red fit).
}
\end{figure}

We obtained the modulation factors by following the method described in
\cite{Lei97}. For the simulated events with 100\% polarized
photons ($\Pi_{s}$=1) the modulation factor can be obtained by fitting the
azimuthal scattering angle distribution with a $\cos 2(\phi-\eta)$ function 
(see Eq.~\ref{eq:pol}). However, before doing this, one needs to take into 
account the ``response'' of the distribution for non-polarized photons. This 
response is obtained by dividing the non-polarized simulated azimuthal 
scattering angle distribution by its average. We divided the 100\% polarized 
azimuthal scattering angle distribution with this response. For the response 
we use, the modulation is on top of an average rate.


\begin{figure}[t]
\plotone{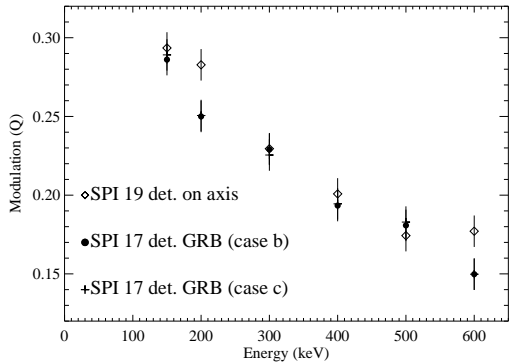}
\caption{\label{fig:mod2}
The simulated modulation factors as a function of energy for three cases; 
diamonds for the hypothetical case of an on-axis GRB (with the same spectra
of GRB~041219) with 19 detectors on, using the actual directions from 
simulations (case b), filled circles and crosses are for the known position of 
\GRB\ with 17 detectors, with case b, and c, respectively (see text for more
explanation on each case). All three cases pixellated.
}
\end{figure}

Fig.~\ref{fig:mod1} shows the azimuthal scattering angle
distribution of 100\% polarized photons at 200 keV as an example.
The amplitude of the modulation with respect to the average gives
the modulation factor. The pixellation reduces the modulation factor
around 20\% (with respect to non-pixellated modulation) at 200 keV.

We ran more simulations with mono-energetic photons at different energies, 
with non-polarized and 100\% polarized photons, with a randomly chosen 
polarization angle of 45$^\circ$. Then we histogrammed the azimuthal scattering
angles using three different methods described earlier. Fig.~\ref{fig:mod2}
shows the distribution of modulation factor as a function of energy for 
different cases. The \wsim 3$^\circ$ off-axis position of the
 GRB, and the reduced number of detectors did not affect the modulation
factors significantly. More importantly, using the energy depositions to 
determine the directions rather than using the actual directions has no effect 
on the modulation factor.


\begin{figure}[t]
\plotone{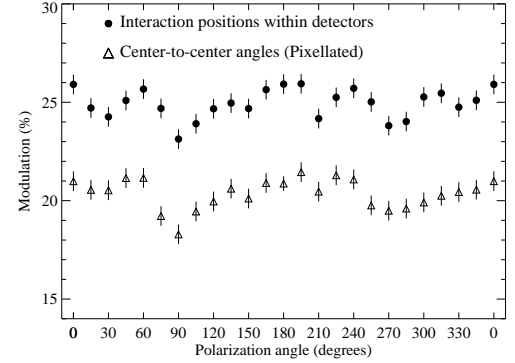}
\caption{\label{fig:modvsan}
Simulated modulation factors for \GRB\ (100--350 keV) as a function of
polarization angle. Interaction positions within detectors are used to obtain
the points shown with circles. The triangles, on the other hand, uses
center-to-center angles. Our best-fit determination of the input spectrum
during R1 is used. 
}
\end{figure}

Finally, by using simulations with the GRB spectrum described above, we
determined the modulation factor for R1 in 100--350 keV band. We ran the
simulations with different polarization angles. The results are shown in
Fig.~\ref{fig:modvsan}. For the pixellated case, $Q$ varies between 18\% and 
21\%.  The average modulation is 20\%. With real interaction positions and 
order, the distribution shows a similar behavior, varying between 23\% and 
26\% with an average of 25\%. Similar to monochromatic tests, the pixellation 
reduces the average modulation factor 20\% for the GRB position and spectrum.

\subsection{Search for systematic effects}

\begin{figure}[t]
\plotone{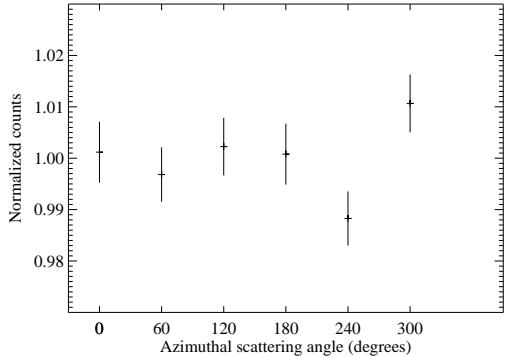}
\caption{\label{fig:blc}
The azimuthal scattering angle distribution of a calibration run with no
mask. The histogram is divided by its average to obtain variations with
respect to the mean. The variations are in the 1\% level.
}
\end{figure}

Since \emph{INTEGRAL} is not rotating, systematic effects not
foreseen by the simulations could alter the azimuthal scattering angle
distribution. Even though we apply corrections to the simulations, there may be
 systematic effects related to ME events that are not discussed in
\cite{Sturner03}. We therefore analyzed some of the data taken at the ground 
calibration tests of SPI (the Bruy\`eres-Le-Ch\^atel dataset, 
\citealt{Attie03}) to search for systematic effects that could affect 
polarization measurements. The best calibration dataset for our purposes is the
case with no mask, on-axis, and using $^{133}$Ba, which resulted in
strong lines at 276.4, 302.9, and 356.0 keV. Unfortunately, after
the runs, it was discovered that the source position was slightly
offset. A Hostaform plastic device inserted in the center of the
plastic anti-coincidence scintillator shadows Detector 0. Due to the
offset, it also casted a shadow on Detectors 2 and 3. We decided to
ignore all interactions between Detector 0 and the surrounding six
nearest neighbor detectors for this analysis. We also excluded the
interactions that involve Detectors 2 and 17, as they are no longer
operational. We only used the interactions for which the total
energy gives the line energy. We did not apply a dead-time
correction, but applied a correction factor to account for differing
detector masses (see \S~\ref{sec:polme}). Then we ran our standard
histogram procedures to obtain the azimuthal scattering angle
distribution. Fig.~\ref{fig:blc} shows this distribution normalized
by its average. The variations are in the 1\% level. Therefore,
excluding interactions in Detector 0, the systematic errors inherent to the
detectors are of the order of 1\%.

\subsection{Chance coincidences}

An important factor in polarization experiments using coincidence events is the
rate of chance coincidences, events in two detectors that occur
within the pre-determined coincidence window. The electronic coincidence window
is 350 ns for SPI, i.e. two events in different detectors that occur
within 350 ns of each other are recorded as a ME. At the peak of the outburst,
the total count rate in singles for the detectors that are not
shadowed is \wsim 400 cts/s. When we eliminate pairs that will not obey our
selection criteria (see \S~\ref{sec:polme}), the maximum singles count rate for
each detector is \wsim 150 cts/s. Therefore the maximum chance coincidence rate
at the peak of the outburst is only 0.008 cts/s per detector pair, which
is negligible for our measurements.

\section{Polarization measurement}\label{sec:polme}

Before analyzing the data, we applied three energy cuts to the multiple events:
the minimum allowed energy for each detector in a pair is 26 keV, the minimum
allowed total energy of a pair is 100 keV, and the maximum allowed total energy
of a pair is 350 keV. The minimum cuts are necessary to ensure that the events
are actual Compton events. They also cut a significant portion of small Compton
scattering angles which contribute less to the modulation factor.  The maximum 
cut is required for two reasons. First, due to low count rates and low 
modulation factors, including the very high energy part does not improve the 
measurement. Second, as discussed in \S \ref{subsec:Q}, as the total energy
increases, the number of events with incorrect azimuthal scattering angles 
increases. To obtain maximum allowed total energy we considered the 
signal/noise ratio of MEs for different energies, their respective modulation 
factors, and finally the fraction of the incorrectly tagged events. And 
finally, we cut all MEs with total energies between 184 keV and 201 keV to 
remove significant number of background photons in the prominent Ge line at
198 keV. 

We then defined two pseudo-detectors (PD) for each detector pair
(i.e. events that scatter from detector 0 to detector 1 is PD1, and
events that scatter from detector 1 to detector 0 is PD2, therefore,
it is different than the Pseudo-detector definition of ISDC). We
only used the nearest neighbors. Even though it is possible to
increase the number of angles by using non-neighbor detectors, the
number of these events are too low to include in the analysis. There
are 64 PDs after the failures of detector  2 and 17.

We separated the light curve in three regions. Region 1 is from the beginning
of the burst to time that the packet-loss problems began. Region 2 and Region 3
are determined using the source and background rates to maximize the source
to background ratio. These regions are denoted as R1, R2, and R3 in
Fig.~\ref{fig:recon}. The analysis is relatively straightforward for R1.
For each PD, the background is determined exactly as determined for singles;
using the first 1000s of the pointing. The rate is again corrected for
dead-time and evolution. The total-background counts are histogrammed into
 6 azimuthal scattering angles. The total number of source counts is 545, and
the total number of background counts is 171.


\begin{figure}[t]
\plotone{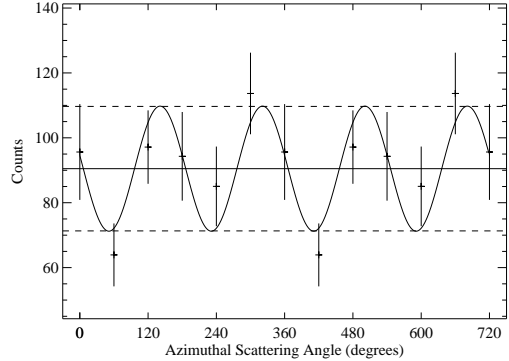}
\caption{\label{fig:polR1}
The azimuthal scattering angle distribution of events in Region 1, and a
$\cos 2(\phi-\eta)$ fit to the data. The solid line is the average
(no polarization) and the dashed lines show the maximum and the minimum
modulation for a 100\% polarization fraction.
}
\end{figure}

The simulated, non-polarized events are corrected for mass and dead-time. The
dead-time for a pair is calculated using both the dead-time due to ACS vetos 
and the detector electronics. We multiplied the number of events in PDs with
${m_{1}+m_{2}}\over{2 \;m_{avg}}$ where $m_{1}$ and $m_{2}$ are
masses of the detectors that form the PD, and $m_{avg}$ is the
average mass of all detectors. After these corrections, we
histogrammed the simulated data exactly as we histogrammed the real
data. To obtain the polarization fraction, we followed the method
described in \citep{Lei97}, and also discussed in Section
\ref{subsec:mgeant}. The resultant distribution and the $\cos
2(\phi-\eta)$ fit is shown in Fig.~\ref{fig:polR1}. The best fit
modulation amplitude is $Q\Pi_{s} = $21.3 $\pm$ 7.6\%, corresponding
to a polarization angle $\eta$ = 48.3$^\circ$ $\pm$ 3.8$^\circ$. The
$\chi^{2}$ for this best fit is 2.69 for 3 degrees of freedom (DOF).
For comparison, the  $\chi^{2}$ for the best fit assuming no
polarization (flat distribution) is 11.00 for 5 DOF. For polarization angles 
$\eta$ \wsim 45$^\circ$, we calculate the polarimetric modulation factor as 
$Q$=21.2 (Fig.~\ref{fig:modvsan}). Correcting the best fit modulation
amplitude for this factor yields a best fit polarization amplitude
of $\Pi_{s}$ = 100 $\pm$ 36\%, providing no upper bound. 

We tried to obtain better constraints by combining Regions 1, 2 and 3. For
Regions 2 and 3, we determined the additional dead-time due to the missing
packets, and corrected the background according to this dead-time. The
remaining of the analysis is the same as Region 1. For the combined case
(R1+R2+R3), the source counts and background counts are 839 and 389
respectively.  Because of the evolution of the GRB spectra, the combined
spectrum is slightly softer than the spectrum of R1. We determined that a Band 
function with $\alpha$=1.15, $\beta$=2.4, $E_{br}$=180 fits the overall 
spectrum well.

\begin{figure}[t]
\plotone{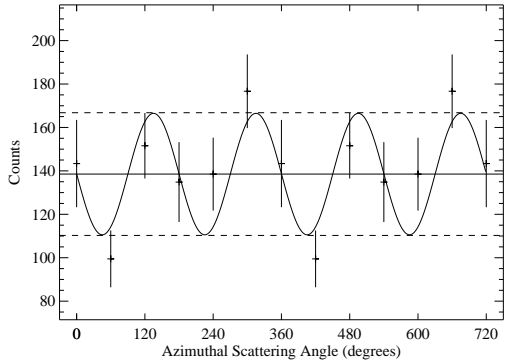}
\caption{\label{fig:polR2}
The azimuthal scattering angle distribution of events in Regions 1, 2 and 3
and a $\cos 2(\phi-\eta)$ fit to the data. The solid line is the average
(no polarization) and the dashed lines show the maximum and the minimum
modulation for a 100\% polarization fraction.
}
\end{figure}

 The azimuthal scattering angle distribution for the combined case is shown in
Fig.~\ref{fig:polR2}. The fit shown yields a modulation amplitude of
$Q\Pi_{s}$=20.2$\pm$6.7\% with a minimum at 45.4$^\circ$ $\pm$ 5.2$^\circ$.
The modulation factor at this angle is 20.4\%, corresponding to 
$\Pi_{s}$ = 99 $\pm$ 33\%. The $\chi^2$ for the $\cos 2(\phi-\eta)$ fit
is 4.68 for 3 degrees of freedom (DOF), whereas the $\chi^{2}$ for the flat
distribution is 15.10 for 5 DOF. Neither of these fits represent the data well,
 as seen in Fig.~\ref{fig:polR2}, and also inferred from the $\chi^2$  values.
Given our measurement uncertainties, and assuming an unpolarized (flat)
distribution, a simple Monte-Carlo simulation yields the chance probability of
 fitting a modulation of this amplitude as 1.01 \%. The best-fit polarization
yields a lower reduced $\chi^{2}$ over the fit assuming no polarization, with
an F-test \citep{Bevington} value of 3.34 with 17.3\% chance probability (over
a flat distribution).

\section{Discussion}\label{sec:discussion}

We have demonstrated techniques to measure polarization
of the prompt \GR\ emission of a GRB in the field-of-view of SPI on
\integral. However, for \GRB, we have not strongly constrained
models for the emission mechanism nor the central engine. ME count rate is not
high enough for statistically significant measurements. For comparison, the
\emph{RHESSI} solar flare polarization measurements utilize approximately  
6500 counts and 16000 counts for two flares \cite{Boggs06}, and we use only 
839 source counts for \GRB.

Another problem is the dependence of the polarization fraction on energy cuts.
The quoted numbers in this work are for the cases with the largest
polarization fractions with the highest F-test values compared to a flat
distribution. However, choosing different energy bands 
for minimum and maximum energies yields lower polarization fractions. For 
example, using 500 keV as the maximum total energy yields a polarization 
fraction of 65 $\pm$ 31\%, with a minimum at 52$^{\circ}$. We obtain 
significantly lower polarization fractions if the 198 keV Ge line is not 
filtered out. If we use energies up to 500 keV and do not cut the 198 keV line,
the polarization fraction is 55 $\pm$ 30\%. The polarization angle may be 
changing with energy, causing a decrease in the overall modulation. 
Unfortunately the statistics are not good enough to test this hypothesis.

Our analysis indicates that systematic effects from the two inactive SPI 
detectors, as determined from pre-flight calibration data, should not 
significantly affect these polarization measurements. We do not have knowledge 
of any further systematic effects in orbit that could affect this 
polarization measurement. Analysis of more GRBs may reveal systematic effects, 
with the potential of distinguishing whether the high modulation we measured 
here was a result of high polarization fraction, a systematic effect, or just 
a chance fluctuation.

The packet loss problem did not play a big role in constraining the
polarization parameters for this GRB. Without any packet loss, there would
have been a sensitivity gain of 15\%, which would not have significantly
affected the upper limit determination, and may have placed a slightly more
stringent lower limit. This exercise showed that SPI has a better chance of
measuring polarization fraction for harder, longer bursts.

\section{Conclusion}

We have used data from SPI on INTEGRAL, an instrument not intended for
polarization studies, and tried to constrain polarization parameters of 
\GRB\, a long and bright GRB in the field of view. The distribution of
 azimuthal scattering angles from \GRB\ is better represented by a polarized
source compared to a non-polarized source, but with  low statistical 
significance. Due to large uncertainties, we have not
strongly constrained models for the emission mechanism, nor for the central
engine. In order to do so, future soft \GR\ missions with polarization 
sensitivity should necessarily aim for an ability to measure polarizations at 
the 5--10\% level, preferably in 3--4 neighboring energy windows, and also 
2--4 intervals spanning a burst's duration. These requirements would render a
\GR\ polarimeter capable of exploring energy-dependent polarization dependence
around the $\nu - F_{\nu}$ peak, and also temporal evolution of both the angle
and degree of polarization. Knowledge of such source characteristics can 
realistically discriminate between some suggested radiation mechanisms and
different model geometries that are currently being contemplated in the GRB
literature.


\acknowledgments E.K. is supported by the European Commission through a FP6 
Marie Curie International Reintegration Grant (INDAM, MIRG-CT-2005-017203). 
E.K. also acknowledges NASA grant NAG5-13142 and partial support of  
T\"UB\.ITAK. M.B. acknowledges NASA grant NNG06GB81G for support. We thank 
Steve Sturner and Chris Schrader at GSFC for providing the BLC data, and their 
helps in MGEANT analysis. We also thank Volker Beckmann for his helps on the 
SPI GRB analysis. We thank Stephane Schanne at SACLAY for clarifying telemetry 
problems, and Pierre Dubath at ISDC for clarifying dead-time routines. We 
acknowledge Mark Kippen and Mark McConnell for the ``GLEPS'' package in MGEANT.
 E.K. thanks Dave Willis and Roland Diehl at MPE for useful discussions.





\end{document}